\documentclass[aps,12pt,preprint,superscriptaddress,showpacs]{revtex4}
\usepackage{psfig}
\usepackage{longtable}
\usepackage{graphics}
\usepackage{graphicx}
\begin{document}
\title{Confirmation of Parity Violation in the Gamma Decay of $^{\rm 180}$Hf$^{\rm m}$}
\author{J.~R.~Stone}
\affiliation{Department of Physics,
University of Oxford, Oxford, OX1 3PU, United Kingdom}
\affiliation {Department of Chemistry and Biochemistry, University of
Maryland, College Park, MD 20742, USA}
\author{G.~Goldring}
\affiliation{The Weizmann Institute of Science, Rehovot, Israel}
\author{N.~J.~Stone}
\affiliation{Department of Physics,
University of Oxford, Oxford, OX1 3PU, United Kingdom}
\affiliation {Department of Physics and Astronomy, University of Tennessee, Knoxville, TN 37996, USA} 
\author{N.~Severijns}
\affiliation{K.U. Leuven, Instituut voor Kern- en Stralingfysica, B-3001 Leuven, Belgium}
\author{M.~Hass}
\affiliation{The Weizmann Institute of Science, Rehovot, Israel}
\author{D.~Zakoucky}
\affiliation{Nuclear Physics Institute, ASCR, 25068 Rez, Czech Republic}
\author{T.~Giles}
\affiliation{AB Department, CERN, CH-1211 Geneva 23, Switzerland}
\author{U.~K\"{o}ster}
\affiliation{Institut Laue Langevin, 6 rue Jules Horowitz, F-38042 Grenoble Cedex 9}
\affiliation{ISOLDE, CERN, CH-1211 Geneva 23, Switzerland}
\author{I.~S.~Kraev}
\affiliation{K.U. Leuven, Instituut voor Kern- en Stralingfysica, B-3001 Leuven, Belgium}
\author{S.~Lakshmi}
\affiliation{The Weizmann Institute of Science, Rehovot, Israel}
\author{M.~Lindroos}
\affiliation{AB Department, CERN, CH-1211 Geneva 23, Switzerland}
\author{F.~Wauters}
\affiliation{K.U. Leuven, Instituut voor Kern- en Stralingfysica, B-3001 Leuven, Belgium}
\date{\today}

\begin{abstract}

This paper reports measurements using the technique of On Line Nuclear Orientation (OLNO) which reexamine the gamma decay of  isomeric $^{\rm 180}$Hf$^{\rm m}$ and specifically the 501 keV 8$^{\rm -}$ -- 6$^{\rm +}$ transition. The irregular admixture of E2 to M2/E3 multipolarity in this transition, deduced from the forward-backward asymmetry of its angular distribution,  has for decades stood as the prime evidence for parity mixing in nuclear states. The experiment, based on ion implantation of the newly developed mass-separated $^{\rm 180}$Hf$^{\rm m}$ beam at ISOLDE, CERN into an iron foil maintained at millikelvin temperatures, produces higher degrees of polarization than were achieved in previous studies of this system.  The value found for the E2/M2 mixing ratio, $\epsilon$ = -0.0324(16)(17), is in close agreement with the previous published average value $\epsilon$ = - 0.030(2), in full confirmation of the presence of the irregular E2 admixture in the 501 keV transition. The temperature dependence of the forward-backward asymmetry has been measured over a more extended range of nuclear polarization than previously possible, giving further evidence for parity mixing of the 8$^{\rm -}$ and 8$^{\rm +}$ levels and the deduced E2/M2 mixing ratio.
\end{abstract}

\maketitle
\newpage
\section{\label{intro}Introduction}
Parity, reflection symmetry in the origin of a co-ordinate system, is one of the fundamental symmetries of physics. Establishing whether parity may be taken as a conserved quantity under a system of forces, or alternatively determining the conditions under which, and the degree to which, parity is not conserved, form basic constraints upon physical theories. The discovery of parity non-conservation (PNC) in the weak interaction was one of the most important discoveries of modern physics. However the extent to which parity is to be considered a conserved quantity in nuclear phenomena remains a challenge to both experiment and theory. Parity mixing in bound nuclear systems is understood as a consequence of weak (parity violating) interaction terms in the nuclear Hamiltonian and precise calculations of this phenomenon are not yet available. 

Of the many experiments aimed at detecting parity non-conservation in nuclear states, one, the measurement of an irregular E2/M2 mixing in the 8$^{\rm -}$ -- 6$^{\rm +}$, highly K-forbidden, 501 keV gamma decay of the 5.47 h isomer of $^{\rm 180}$Hf, stands out. It was first observed in gamma ray circular polarization experiments \cite{jen70,lip71}. Using the technique of low temperature nuclear orientation (LTNO), the reported result is an apparently well established effect, a mixing ratio $\epsilon$(E2/M2) = -0.030(2), of a magnitude more than fifteen times the experimental error \cite{kra71a,kra71b,kra72,kra75}. Other LTNO measurements are listed in \cite{kra86}. Two other statistically significant PNC results, of much smaller experimental effects, on states in $^{\rm 175}$Lu \cite{kup74}  and  $^{\rm 181}$Ta \cite{lip72}, have been reported using circular polarization technique. On no other bound nuclear systems, including the recent work on $^{\rm 93}$Tc$^{\rm m}$ \cite{nar05}, do the latest published reports claim to find any effect deviating by more than two standard deviations from parity conservation \cite{kra86,ade85}. The result on the  $^{\rm 180}$Hf$^{\rm m}$ isomeric decay, based on experimental evidence of both angular distribution studies from nuclei polarized at millikelvin temperatures and circular polarization studies (see \cite{kra72} and references therein), dating from the 1970's, stands today as the strongest evidence of the level to which nuclear states cannot be taken as eigenstates of parity.

It is important to examine the validity of this significant result. This paper describes a remeasurement of the evidence for parity non-conservation effects in the $^{\rm 180}$Hf isomeric decay, using the techniques of on-line nuclear orientation available today. Developments of technique over 30 years have given access to different source making methods, leading to higher degrees of polarization and the ability to make continuous observations over a period of days rather than successive measurements on a series of decaying samples.

The paper starts with description of previous work, a brief introduction to the necessary formalism and the justification of a new study in Sec.~\ref{form}. This is followed (Sec.~\ref{exp}) by the detailed description of the new experiment and the analysis of the gamma ray spectral data. Comparison with theoretical calculation of the observed effects requires discussion of aspects of the hyperfine interaction and of the angular distribution coefficients, given in Sec.~\ref{param}. Sec.~\ref{further} includes discussion of the quality of implantation and aspects of thermometry, leading to evaluation of the parity violating E2/M2 mixing ratio in the 501 keV transition. A brief discussion of the final results and the need for, but difficulty of, its interpretation in nuclear theory, is given in Sec.~\ref{disc}.

\section{\label{form}Formalism, previous experiments and technical developments.}
\begin{figure}
 \centerline{\psfig{file=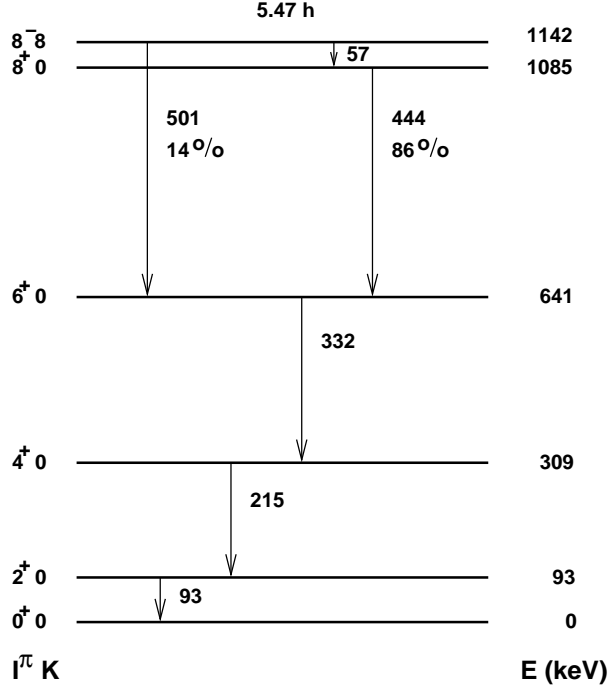,width=8cm}}
 \caption{\label{fig1}Decay of the 5.47 h  $^{\rm 180}$Hf$^{\rm m}$}
 \end{figure}

Full descriptions of the formalism of the nuclear orientation experimental method and of the previous measurements on  $^{\rm 180}$Hf$^{\rm m}$ are given in Refs.~\cite{kra71a,kra71b,kra72,kra75,kra86}. Here only the essential details relevant to the present work and its analysis are presented. Fig~\ref{fig1}
 shows a partial level scheme of $^{\rm 180}$Hf with the 5.47 h, I$^{\rm \pi}$= 8$^{\rm -}$, K=8, isomer decaying to the 8$^{\rm +}$ and 6$^{\rm +}$ levels of the K=0 ground state band. In a nuclear orientation experiment the radioactive sample is polarized by cooling to millikelvin temperatures in an environment in which the nuclei experience a magnetic field sufficient to produce a high and controllable degree of nuclear polarization. The gamma radiation from such a polarized source exhibits a strongly anisotropic angular distribution. If the nuclear states between which the gamma decay occurs are eigenstates of parity, the theoretical distribution contains only even terms in cosine theta, measured from the axis of polarization, and is given by the expression \cite{kra75}
\begin{equation}
W(\theta,T)=1+\sum_{\lambda \, \rm even}fB_\lambda(T) U_\lambda A_\lambda Q_\lambda P_\lambda(\cos\theta)
\label{w}
\end{equation}
where B$_{\rm \lambda}$(T) are the temperature dependent orientation parameters, describing the polarization of the parent nuclei, U$_{\rm \lambda}$ are angular momentum coupling constants describing the change in the polarization when the decay passes through a sequence of unobserved transitions between states before the detected gamma emitting state is reached, A$_{\rm \lambda}$ are angular momentum coupling constants describing the multipolarity of the observed transition, Q$_{\rm \lambda}$ are solid angle correction factors for finite detector size and P$_{\rm \lambda}$ are associated Legendre polynomials. The factor $f$, the fraction of nuclei in sites which experience the hyperfine field, is particular to implanted sources and is fully discussed in Sec.~\ref{fra}. It is usual to place detectors at 0$^{\rm o}$, 90$^{\rm o}$ and 180$^{\rm o}$ to the polarization axis and to describe the orientation in terms of the anisotropy $a$(T) given by
\begin{equation}
a(T) = \frac{W(0^{\rm o},T)}{W(90^{\rm o},T)} - 1.
\label{anis} 
\end{equation}
The normalized intensities at 0$^{\rm o}$ and 180$^{\rm o}$ are the same under the assumption of parity conservation.

If either of the nuclear states between which the gamma decay occurs are not eigenstates of parity, odd $\lambda$ terms in the distribution, of the same form as Eq.~\ref{w} are introduced. These terms are seen most clearly in the asymmetry $\mathcal{A}$(T), defined in 
\cite{kra71a,kra71b,kra72} as
\begin{eqnarray}
\mathcal{A}(T)  =  2\frac{W(0^{\rm o},T)- W(180^{\rm o},T)}{W(0^{\rm o},T)  +  W(180^{\rm o},T)} \\ 
=\frac{2\sum\limits_{\lambda \, \rm odd}fB_\lambda(T) U_\lambda A_\lambda Q_\lambda P_\lambda(\cos\theta)}{1  +  \sum\limits_{\lambda \, \rm even}fB_\lambda(T) U_\lambda A_\lambda Q_\lambda P_\lambda(\cos\theta)}.
\label{asymmetry}
\end{eqnarray}
The previous work sought evidence for parity non-conserving admixtures in both the 57 keV 8$^{\rm -}$ --~8$^{\rm +}$ transition and the 501 keV 8$^{\rm -}$ --~6$^{\rm +}$ transition by measuring the asymmetry between the gamma intensity measured at zero degrees and 180 degrees to the axis of polarization of samples of $^{\rm 180}$Hf$^{\rm m}$ in (Hf$_{\rm x}$Zr$_{\rm 1-x}$)Fe$_{\rm 2}$ alloys
cooled to millikelvin temperatures \cite{kra71a,kra71b,kra72,kra75}. For the 57 keV transition the reported gamma ray asymmetry was small, consistent with zero to one part in a thousand. However the 501 keV transition consistently showed asymmetry of close to 1.5\% at a temperature of about 20 mK, more than ten times the experimental error.

Two, related, major technical developments over the thirty years since the last reported experiments on $^{\rm 180}$Hf$^{\rm m}$ decay make an experimental re-examination of the case timely. The first of these is the development of the on-line nuclear orientation technique in which a beam of radioactive ions is implanted into a ferromagnetic metal host foil cooled to millikelvin temperatures in a $^{\rm 3}$He/$^{\rm 4}$He dilution refrigerator with side access for the beam \cite{ede90}. This method allows continuous measurement at a controllable range of temperature down to below 15 mK with a uniform source strength. In contrast, the previous experiments involved a series of samples cooled by contact with a demagnetized paramagnetic salt, which thereafter steadily warmed giving a slowly changing degree of nuclear polarization and each successive sample decaying with the 5.47 h half-life giving rise to variable dead-time and electronic pile-up correction. The second is the development of a means to produce a beam of ionized Hf, a highly refractory element, from an isotope separator.

As well as gaining access to better controlled temperatures and a steadier source strength, the use of ion implantation allows use of pure iron, rather than the (Hf$_{\rm x}$Zr$_{\rm 1-x}$)Fe$_{\rm 2}$ alloy, in which to polarize the nuclei. The hyperfine field at Hf in iron is close to 70 T (see below), more than 3 times stronger that the $\sim$ 20 T present in the alloy. This means that the hafnium nuclei can be more fully, indeed almost completely, polarized at attainable temperatures.

In view of these basic improvements of available technique a new experiment on the anisotropic emission of gamma radiation from oriented $^{\rm 180}$Hf$^{\rm m}$ has been carried out. 

\section{\label{exp}The experiment}
\subsection{\label{outline}Outline}
\begin{figure}
\centerline{\psfig{file=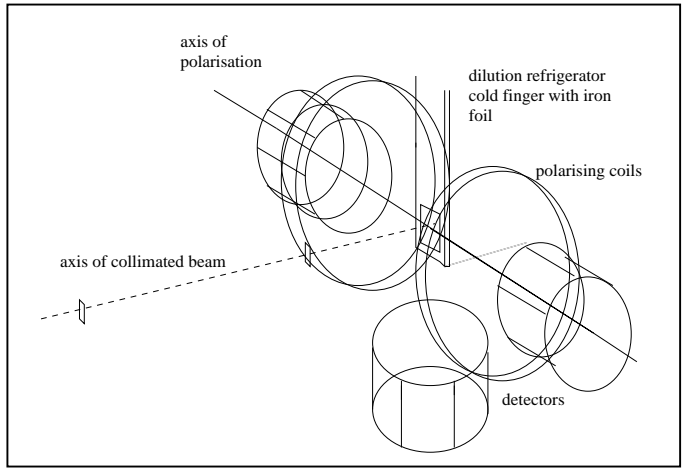,width=9cm}}
\caption{\label{fig2}Schematic experimental set-up.}
\end{figure}

The experiment was performed at the ISOLDE isotope separator facility, CERN, using the NICOLE on-line nuclear orientation dilution refrigerator system \cite{ede90}. The 1.4 GeV protons were incident upon a mixed Ta/W metal foil target and the hafnium atoms produced were transfered to the ion source using fluorine, added to the plasma support gas as CF$_{\rm 4}$. The most intense emerging hafnium fluoride molecular ion beam, used for the experiment, was HfF$_{\rm 3}^+$ \cite{kos07}. The Hf orientation sample was prepared by accelerating the ions to 60 keV and, after mass separation, impinging them on the surface of a pure (99.99\%) iron foil soldered to the copper cold finger of the dilution refrigerator, which is perpendicular to the separator ion beam (see Fig.~\ref{fig2}). At impact the molecular ions disintegrated and the hafnium nuclei were implanted into the iron lattice, producing a source entirely free of contaminant activity. The F nuclei also enter the iron but come to rest in a region well removed from the Hf nuclei. The total implantation dose was approximately 4 x 10$^{\rm 11}$ ions into a region of $\approx$3 mm in diameter giving a maximum local concentration of Hf in the iron foil below 10$^{\rm -3}$ atomic percent. The quality of the implantation, as indicated by the fraction of hafnium nuclei stopping in substitutional lattice sites in the iron, is discussed below. A second iron foil, containing diffused $^{\rm 57}$Co activity for which all orientation properties are known, was soldered to the back of the cold finger to act as a nuclear orientation thermometer \cite{mar86}. During the experiment the iron foil sample was magnetized to saturation using a Helmholtz pair of polarizing coils which produced a magnetic field of 0.5 T applied in the plane of the foil and at right angles to the beam axis. The direction of the field, which determines the sense of polarization of the radioactive nuclei, could be reversed by changing the direction of the current through the coils, a procedure which took about 10 minutes.

 Gamma radiation from both activities was detected in three large intrinsic germanium detectors, two placed on the field axis on opposite sides of the cryostat, at 0$^{\rm o}$ and 180$^{\rm o}$ to the axis of polarization (depending upon the field direction) and one at 90$^{\rm o}$ to the axis of polarization below the cryostat. The spectra were accumulated in files of 300 s duration throughout the experiment, pulses from a regular pulser being introduced to the pre-amplifiers to allow correction for pile-up in detectors and dead time in the electronics.

\begin{figure}
\centerline{\psfig{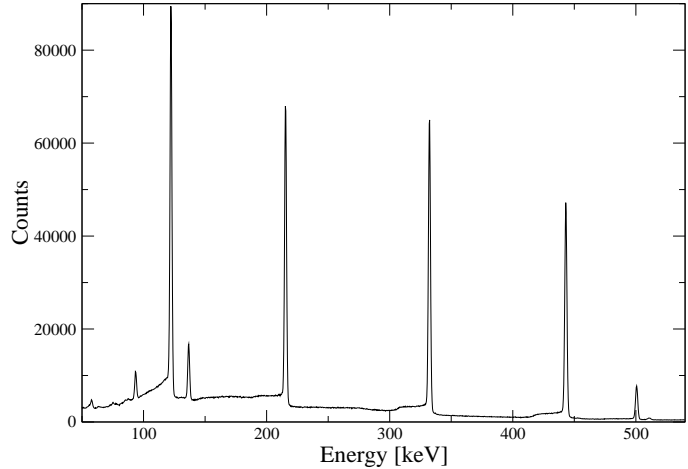}}
\caption{\label{fig3}Gamma-ray spectrum showing $^{\rm 180}$Hf$^{\rm m}$ transitions at 501, 444, 332 and 215 keV and $^{\rm 57}$Co transitions at 137 and 122 keV, with no Pb absorber present.}
\end{figure}

The gamma ray spectrum (see Fig.~\ref{fig3}) contained six strong, fully resolved, transitions, at 501 keV, 444 keV, 332 keV and 215 keV from  $^{\rm 180}$Hf$^{\rm m}$ (the 57 keV and 93 keV transitions were strongly absorbed in cold finger and cryostat), and at 137 keV and 122 keV from the $^{\rm 57}$Co thermometer. To obtain counts for each transition detected in each 300 s file, simple windows were set over the peaks, taking care to set them wide enough so that any small changes in gain of the system did not result in loss of counts at any time during the experiment. Background to each peak was found by setting additional windows on the spectra above and below the peaks and making a linear interpolation to determine the background count in each peak window.

The sample was cooled to achieve high degrees of nuclear polarization and the polarizing field reversed at intervals when the refrigerator was operating steadily and the implanted ion beam was stable. At each reversal the change in gamma intensity recorded in each of the two axial detectors was measured in all transitions. For transitions in which parity is conserved there should be no change in intensity since any observed change is proportional to the parity non-conserving transition amplitude. The principal objective of the experiment was to observe the temperature dependence of the parity non-conserving amplitude reported in the 501 keV transition over a wider range of temperature, and hence to higher degrees of nuclear polarization, than had been accessible to the previous studies (see Fig.~\ref{fig4}).

After the beam was first introduced to the refrigerator, the  $^{\rm 180}$Hf$^{\rm m}$ activity was allowed to accumulate for about 5 hours with the temperature close to 1 K and the iron foil unpolarized. The polarizing field was then applied and a sequence of `warm' reference spectra were taken with negligibly small nuclear orientation. As the source strength approached its asymptotic value, the sample was cooled to about 13.5 mK, the lowest accessible temperature with the beam present. Just after this, interruption of the beam from ISOLDE reduced the heat input, allowing further cooling to the refrigerator base temperature, 7.6 mK. This complete sequence is referred to as the initial cool-down. Shortly after the beam had returned, the first of a series of eight reversals was carried out. For each reversal the temperature was measured using nuclear orientation thermometry as described later. The first six reversals were done with the implantation beam incident on the sample. They were accordingly at temperatures which reflected balance between beam and radiation heating to the cold finger, plus heating from absorption of the gamma activity in the source, and the cooling power of the refrigerator. 
The temperatures ranged between 15.6 and 25 mK, increasing slowly with time as the beam intensity and source strength grew. Later the mixing chamber of the refrigerator was heated to give a temperature close to 60 mK, still with implanted beam, and a sixth reversal took place at this relatively high temperature. Finally at the end of the experiment, when implantation was stopped, the sample cooled to the refrigerator base temperature of 7.6 mK during the decay of  the activity and the seventh, lowest temperature, reversal was made. An eighth, `dummy', reversal was made after the sample was warmed to about 1 K when no orientation was present.
\subsection{\label{first} A first survey of the results.}
For the 501 keV transition to show appreciable asymmetry, assuming that the other transitions in the $^{\rm 180}$Hf$^{\rm m}$ decay are symmetric, a simple way to show the effect is to consider the ratio of the peak counts $N$ in the (asymmetric) 501 transition to those in another (symmetric) transition, for example that at 444 keV, or combination of symmetric transitions. As is shown in Appendix A (for the case in which the sum of the 444 keV and 332 keV counts is taken as the symmetric `norm') the change in such a ratio, measured in a single detector on the axis of orientation, when the direction of nuclear polarization is reversed, is directly proportional to the asymmetry of the 501 keV transition.
\begin{figure}
\begin{minipage}[t][8cm][t]{10cm}
\includegraphics[width=9cm,height=6cm]{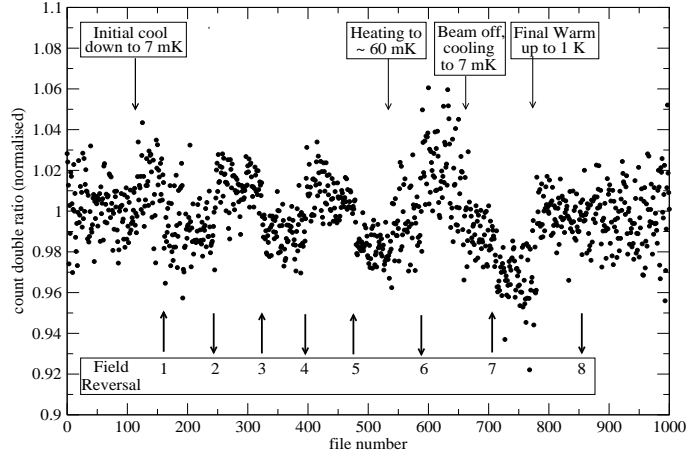}
\end{minipage}
\begin{minipage}{10cm}
\includegraphics[width=9cm,height=6cm]{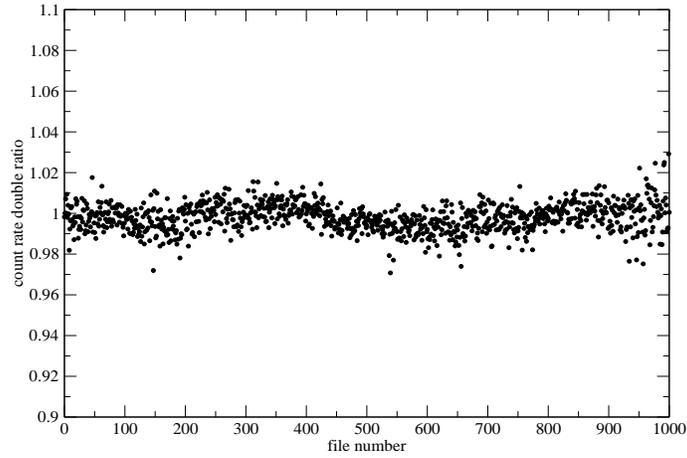}
\caption{\label{fig4} Upper panel: The double ratio $\mathcal{R}$ of counts in the 501 keV and 444 keV peaks measured in the Left(L) and Right(R) on-axis detectors for all files of the experiment, normalized to the average pre-cool-down value of $\mathcal{R}$ (files 73-86). Lower panel: As upper, but for the double ratio of the 332 keV and 444 keV peak counts, which should exhibit a null effect. For discussion, see text.} 
\end{minipage}
\end{figure}

Fig.~\ref{fig4} (top panel) shows the measured double ratio
\begin{equation}
\mathcal{R} = \frac{N(501,L)/N(444,L)}{N(501,R)/N(444,R)}
\end{equation}
 for each file during the experiment. Here L and R refer to the two detectors on the axis of nuclear polarization. When the applied field, and hence the direction of nuclear polarization, is reversed, the asymmetry, and with it any difference in this ratio from unity, must change sign. Fig.~\ref{fig4} (bottom panel) shows the same ratio, but for the 332 keV transition to the 444 keV transition. The file numbers at which field reversals were made are indicated by the sequence of upward and downward arrows. In both panels the scatter of the data indicates the statistical error. It is immediately apparent that the data in the upper panel show variation about unity which reverses at every field reversal except the eighth. This last reversal took place above 1 K, with zero nuclear polarization, so a null effect is expected. By contrast, the data in the lower panel exhibit no such field direction dependence, as is expected for transitions between good eigenstates of parity. It is however also clear that the data in the upper panel show, in addition, some variation from unity when the temperature of the dilution refrigerator, i.e. the degree of nuclear polarization and hence the count rate in the different detectors, is changing rapidly. Cooling and heating took place over relatively narrow ranges of file numbers whose centres are indicated in the top of the figure. The apparent asymmetry suggested by these additional changes is due to different response of the detectors to strongly changing count rate. Such spurious effects have been fully eliminated in the more comprehensive analysis presented in the following sections. 

\subsection{\label{detail}Detailed description of the experiment and spectral analysis of the field reversals.}
In this section the procedure is described in sequence and in considerable detail, allowing full discussion of the several stages of the experiment.

\subsubsection{\label{sel}Selection of data files for analysis.}
The sample temperature at any time is determined by a balance between a combination of heating and cooling processes. During the early part of the experiment many short interruptions of the beam took place, during which the sample temperature fell rapidly over even just a few 300 s data files, recovering to its pre-interruption value within a few files of the return of the beam. All files involved with such variations of temperature were set aside, as were others during which the dilution refrigerator was being filled with cryogens, which can also somewhat perturb its operation. The files used in the analysis were free of any detectable outside perturbations. During each magnetic field reversal the current in the polarizing coils was first reduced over a period of about 300 s then the leads to the power supply were manually interchanged and the current increased over a similar period. This procedure resulted in small changes in the refrigerator temperature as detected in all gamma transitions, caused by eddy current heating in the cold finger holding the sample and the mixing chamber. The initial temperature was recovered in a further three or four files. Data files taken during these periods of heating and recovery were also discarded in the analysis presented below. 


\subsubsection{\label{meas}The measured quantities.}
Since the source strength during implantation is a variable quantity, all anisotropy and asymmetry measurements are in the form of ratios of counts in detectors at different angles to the polarizing field. Furthermore, to eliminate effects of variable dead time and pile up during different counting periods, each raw peak count is divided by the pulser peak count for the same file and detector, referred to as the pulser normalized count. The quantity W$_{\rm exp}$($\theta)$ is the ratio of the pulser normalized count in a specific gamma peak in a detector at angle $\theta$ to the polarization axis, to the pulser normalized counts in that peak from an unpolarized sample, the `warm' counts. The measured value of anisotropy $a(T)$ is given by the ratio [W$_{\rm exp}$(0$^{\rm o}$)/W$_{\rm exp}$(90$^{\rm o}$) - 1]  from which temperatures are deduced and the measured asymmetry $\mathcal A$(T) is given by 2[W$_{\rm exp}$(0$^{\rm o}$) - W$_{\rm exp}$(180$^{\rm o}$)]/[W$_{\rm exp}$(0$^{\rm o}$) + W$_{\rm exp}$(180$^{\rm o}$)], sensitive to the degree of parity non-conservation. Both have been defined above.
\subsubsection{\label{abs}Introduction of absorbers.}
The initial cool-down, combined with the later cool-down after the implanted beam was finally stopped, provided data allowing comparison of the anisotropies measured on the $^{\rm 180}$Hf$^{\rm m}$ and $^{\rm 57}$Co transitions over a wide temperature range. Since temperatures can be deduced from the observations on $^{\rm 57}$Co this allowed calibration of the $^{\rm 180}$Hf$^{\rm m}$ transitions as secondary thermometers so that counts from the $^{\rm 57}$Co transitions were no longer required for temperature measurement. Lead absorbers of equal thickness, giving rise to attenuation factors (measured to be equal to $\pm $1\%) for a given gamma energy in all detectors, were secured over each detector face to reduce the total counting rates. The close equality of the attenuation factors allow use of the pulser normalized `warm' count \textit{ratios} in analysis of the ratios measured with absorbers present.

\subsubsection{\label{rev16}Field reversals 1--6.}
Following the initial cool-down, once the beam returned and the temperature had reached equilibrium, the first field reversal was carried out.
Further reversals were made at intervals of about 7 hours, the time between them being determined largely by the requirement of steady beam and good temperature stability. The results of all field reversals and the conditions under which they were performed are summarised in Table~\ref{tab1}. During this sequence the temperatures deduced from the $^{\rm 180}$Hf$^{\rm m}$ transitions rose slowly from 15.6 mK to close to 25 mK as the sample strength grew, reflecting a slow increase in yield from the ion source.  Reversals 2--4 were carried out in the same experimental set-up. For reversal 5 the pulser, which produced some distortion of the gamma transition peak shapes through a small undershoot in the amplifier base line, was removed to check that its presence did not affect the measurements. Lack of pulser normalization for this reversal (and for reversal 7) meant that rather than being able to evaluate the asymmetry for each transition separately, only the difference of asymmetry between a pair of transitions can be extracted. This is discussed fully in Sec.~\ref{ext} and in Appendix A.

The pulser was reintroduced before reversal 6, for which the refrigerator temperature was raised to close to 60 mK by supplying heat to the mixing chamber. For this reversal $^{\rm 57}$Co thermometry was needed, thus the lead absorbers were removed.

\subsubsection{\label{after}After-beam cool-down, low temperature field reversal 7 and final warm-up.}
The $^{\rm 180}$Hf$^{\rm m}$ beam was stopped shortly after reversal 6 and the dilution refrigerator cooled fairly quickly to its base temperature as the sample slowly decayed. The absence of lead absorbers during the cooling allowed use of the $^{\rm 57}$Co thermometer. Reversal 7 was carried out when the temperature had reached its equilibrium value of 7.6 mK and was made with lead absorbers present but without use of the pulsers. When this reversal was complete the refrigerator was warmed to above 1 K and a `dummy' reversal was performed whilst additional `warm' reference files were recorded for approximately 24 hours. The ratios of pulser normalized gamma peak counts from the initial and final `warm' files in all detectors were unity to within 0.6\%, demonstrating that there had been no significant drift. It should be made clear that this 0.6\% variation over the course of the experiment does not affect the asymmetries derived from the data. This follows since upon each reversal the asymmetry is the immediate change in the ratio of the normalized count rates in the 501 keV transition in the 0$^{\rm 0}$ and 180$^{\rm 0}$ detectors, so normalization to warm counts cancels. In interpreting the asymmetry using Eqn.~\ref{asymmetry} the numerator is directly given by the reversal data. The denominator (the sum of the anisotropies) does depend on the `warm' counts. However, in the even terms, a small change, such as the 0.6\% referred to, simply adds to the anisotropy in one direction and subtracts from the other, leaving their sum unchanged. 

\section{\label{param}Parameters used in calculation of the angular distributions.}
\subsection{\label{mm}The magnetic moment of $^{\rm 180}$Hf$^{\rm m}$.}
The 8$^{\rm -}$ isomeric state of $^{\rm 180}$Hf$^{\rm m}$ is expected to be a rather pure two-quasi-particle state, very similar to the 8$^{\rm -}$ isomers found in other Hf isotopes. Only one of these, $^{\rm 172}$Hf$^{\rm m}$ has a measured magnetic dipole moment, $\mu$($^{\rm 172}$Hf$^{\rm m}$, 8$^{\rm -}$) = 7.86(5) n.m. \cite{wal80}. This moment, however, is in excellent agreement with predictions for a pure two-proton excitation comprising g$_{\rm 7/2+}$ and h$_{\rm 9/2-}$ protons. Taking the g-factors for these two protons from neighboring isotopes, simple combination gives prediction for the moment of the 8$^{\rm -}$ isomer between 7.7 and 7.9 n.m. Although there are many measurements of the g$_{\rm 7/2+}$ single quasi-particle state in the vicinity, and they show only a small variation with neutron number, there is only one accurate measurement of the h$_{\rm 9/2-}$ state  (in $^{\rm 181}$Ta) \cite{sto05}, thus any changes in the magnetic moment of the 8$^{\rm -}$ Hf isomers with addition of neutrons cannot be argued from experimental evidence. However they are likely to be small.

Low precision measurements for the magnetic moment in $^{\rm 180}$Hf$^{\rm m}$  have been reported \cite{kra71a,kra72}, based on the hyperfine splitting extracted from LTNO in the Hf alloy samples used in that work (see below), giving results of close to 9 n.m. with errors of 1 n.m. These appear high compared with estimates based on the nuclear structure of the isomer.   

\subsection{\label{hf}The hyperfine field acting at $^{\rm 180}$Hf$^{\rm m}$ in iron.}
The magnetic hyperfine interaction at a substitutional site in cubic iron is a well defined quantity which involves the product of the magnetic dipole moment of the isotope concerned and the hyperfine field acting at that site. Until recently there were no accurate measurements of the magnetic dipole moment of any radioactive isotope of hafnium \cite{sto05} and thus only inaccurate values of the hyperfine field could be obtained \cite{rao}. However, recently the magnetic moment of $^{\rm 175}$Hf was measured to be - 0.677(9) n.m. \cite{nie02}  and nuclear magnetic resonance of $^{\rm 175}$Hf oriented in iron was reported at a frequency of 139.0(1) MHz \cite{mut04}. These results yield the hyperfine field for Hf in iron as - 67.5(9) T. 

Estimate of the interaction strength for $^{\rm 180}$Hf$^{\rm m}$ in iron can be obtained from published experimental results only through taking the hyperfine interaction expressed as a temperature, T$_{\rm int}$= $\mu$B$_{\rm hf}$/Ik = - 8.2(2) mK determined for $^{\rm 180}$Hf$^{\rm m}$ in the compound (Hf$_{\rm 0.1}$Zr$_{\rm 0.9}$)Fe$_{\rm 2}$ \cite{kra75} and the ratio 3.9(4) of the strengths determined from M\"{o}ssbauer studies on the 93.3 keV 2$^+$ state in iron, i.e. T$_{\rm int}$ = - 7.6(1) mK, and in the compound (Hf$_{\rm 0.1}$Zr$_{\rm 0.9}$)Fe$_{\rm 2}$, T$_{\rm int}$ = - 1.93(17) mK \cite{kor71}. These values give the interaction in iron as T$_{\rm int}$ = - 32.4(30) mK. 

\begin{figure}
\centerline{\psfig{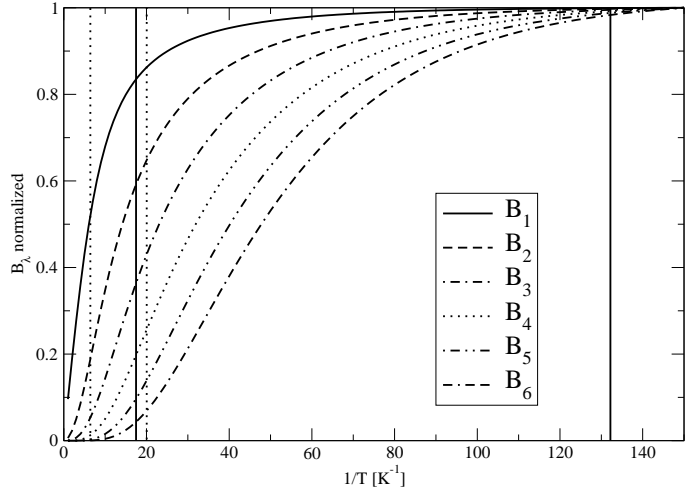}}
\caption{\label{fig5}Calculated B$_{\lambda}$ coefficients vs 1/T for hyperfine interaction 25.9 mK, each normalized to its saturation value at T = 0. The vertical dotted lines show the range of B$_\lambda$ accessed by previous measurements \protect\cite{kra71a,kra71b,kra72,kra75}; full lines indicate the range covered by present work.}
\end{figure}

However, taken with the new, accurate, result for the hyperfine field, such an interaction would predict a magnetic moment of 10.4(10) n.m. for $^{\rm 180}$Hf$^{\rm m}$, far larger than the measured moment in $^{\rm 172}$Hf$^{\rm m}$ and above any reasonable increase due to additional neutrons in $^{\rm 180}$Hf$^{\rm m}$. This is considered to be too high and therefore an estimated interaction, T$_{\rm int}$ = - 25.9(13) mK, based on the measured hyperfine field and a magnetic moment of 8.4(5) n.m. consistent with, but somewhat larger than, the measured value in $^{\rm 172}$Hf$^{\rm m}$ has been used to calculate the orientation parameters B$_\lambda$ for $^{\rm 180}$Hf$^{\rm m}$ in iron. The results for $\lambda$ = 1 - 6 are plotted in Fig.~\ref{fig5}.
 as a function of inverse sample temperature 1/T. In the figure, each B$_\lambda$ is normalized to its saturation value. The range of the orientation parameters at which field reversals were carried out in this work is indicated in the figure and compared with the range accessed by the experiments of Refs.~\cite{kra71a,kra71b,kra72,kra75} showing the increase in the degree of polarization and the close approach to full saturation achieved in the present experiment. The very high degree of polarization achieved has the important consequence that, with uncertainty in the hyperfine interaction estimated to be 6\%, the uncertainty in the extracted irregular parity admixture $\epsilon$ is less than 1\%, small compared with the statistical errors.

\subsection{\label{spin}The nuclear spin-lattice relaxation time of $^{\rm 180}$Hf$^{\rm m}$   and $^{\rm 57}$Co in iron.}
To assure full thermal equilibrium between the iron lattice, taken here to be at constant temperature, and the implanted hafnium nuclei prior to their decay it is necessary that the hafnium nuclear spin-lattice relaxation time T$_{\rm 1}$ be short compared to the nuclear lifetime. For nuclei in metals, relaxation is via the conduction electrons and can be described in terms of the Korringa constant C$_{\rm k}$ = T$_{\rm 1}$T, where T is the absolute temperature.  For a given system in iron the value of C$_{\rm k}$ can be estimated from the empirical relation C$_{\rm k}$T$_{\rm int}^{\rm 2}$ = $\sim$ 1.4 x 10$^{\rm -4}$ sK$^{\rm 3}$ \cite{rei67} where T$_{\rm int}$ is the nuclear interaction strength, T$_{\rm int}$ = - 25.9 mK for $^{\rm 180}$Hf$^{\rm m}$. This yields C$_{\rm k}$ $\approx$ 0.21 sK and an estimated relaxation time of 2.0 s at 100 mK. At lower temperatures the simple inverse relation to lattice temperature breaks down and T$_{\rm 1}$ reaches a maximum value approximately given by 3.3 C$_{\rm k}$/(I + $\frac{1}{2}$)T$_{\rm int}$ \cite{shaw}, estimated for $^{\rm 180}$Hf$^{\rm m}$ to be $\approx$ 3 s, far shorter than the lifetime of 5.47 h, thus clearly satisfying the requirement for thermal equilibrium.

As described below, knowledge of the fraction, $f$, of hafnium nuclei in substitutional lattice sites of the iron foil, is necessary both for analysis of the measured asymmetry of the 501 keV transition and also to calibrate the hafnium gamma transition anisotropies for use as thermometers for part of the experiment. The determination of the fraction is done by comparing the anisotropies of the $^{\rm 180}$Hf$^{\rm m}$ and $^{\rm 57}$Co gamma transitions over as wide a range of temperatures as possible.  When the lattice temperature is changing, for example during cool-down, for such comparison to be valid it is necessary that for both the hafnium and cobalt isotopes the spin-lattice relaxation times be substantially shorter than the time of measurement for one file.
For $^{\rm 180}$Hf$^{\rm m}$ this is already established. For $^{\rm 57}$Co T$_{\rm int}$ is 14.2 mK and the Korringa constant is C$_{\rm k}$ = 0.4 sK \cite{lau74} so that the relaxation time varies between 4s at 100 mK and a maximum of 23 s well below 14 mK, again both much shorter than the file measurement time of 300 s. Thus at all temperatures in this experiment the two isotopes can be taken as being in good thermal equilibrium with the iron lattice and with each other.    

\subsection{\label{ang}The gamma transition angular distribution coefficients.}
Another measured parameter required in the angular distribution calculation is the E3/M2 multipole mixing ratio in the 501 keV transition. The value $\delta$(E3/M2) = + 5.3(3) given in Ref.~\cite{kra71b} has been taken. The uncertainty in this parameter produces an error in the final result for the irregular E2/M2 mixing ratio $\epsilon$ again much smaller than the statistical errors. The 57 keV 8$^{\rm -}$ --~8$^{\rm +}$ transition was taken as pure electric dipole \cite{kra72}. All other transitions are of pure electric quadrupole multipolarity. The intensities of the 501 keV and the (unobserved) 57 keV -- 444 keV decay paths feeding the 641 keV 6$^+$ level were taken as 14\% and 86\% of the total respectively \cite{wu03}. 

In Table~\ref{tab2} the calculated values of the even term angular distribution parameters U$_\lambda$A$_\lambda$Q$_\lambda$ for each analysed transition in the decay of $^{\rm 180}$Hf$^{\rm m}$ are given. The parity non-conserving terms for the 501 keV transition are given to first order in the irregular mixing ratio $\epsilon$ by \cite{kra71b}
\begin{equation}
A_\lambda = \frac{2\epsilon}{1+\delta^2}[F_\lambda(2288)+\delta F_\lambda(2388)].
\label{al}
\end{equation}
For $\epsilon$ = - 0.030 and $\delta$ = + 5.3 the values are: U$_{\rm 1}$A$_{\rm 1}$Q$_{\rm 1}$ = - 0.0045, U$_{\rm 3}$A$_{\rm 3}$Q$_{\rm 3}$ = 0.0016 and U$_{\rm 5}$A$_{\rm 5}$Q$_{\rm 5}$ = 0.0019. Here the value of epsilon from \cite{kra75} is taken; the variation of the odd A$_\lambda$ with epsilon is given by Eqn.~\ref{al} and illustrated in Fig.~\ref{fig8}.
\begingroup 
\squeezetable 
\begin{table} 
\caption{\label{tab1}Analysis of both $^{\rm 180}$Hf$^{\rm m}$ and $^{\rm 57}$Co gamma ray data to give asymmetry values for all field reversals.}
\begin{ruledtabular}
\begin{tabular}{cccccc|clrrrrr}
Reversal  & $^{\rm 180}$Hf$^{\rm m}$  &  Pb  & Field & Pulser & T & \multicolumn{6}{c}{ Asymmetry $\mathcal{A}$(T) (\%)}  \\
\cline{7-12}
number    &     Beam   &     & Direction &       &  (mK) &   122 keV  &  137 keV  &  215 keV  &  332 keV  &  444 keV  &  501 keV  \\
\hline
 1  &  ON  &  OFF &  L $-$ R &  ON  &15.6(2)  & +0.08(11)  &  -0.26(47)  &  +0.12(15)  &  -0.07(14)  &  +0.21(15)  &  -1.09(32) \\
 2  &  ON  &  ON  &  R $-$ L &  ON  &19.2(13) &\multicolumn{2}{c}{ lead absorber present}& -0.27(19)  &   -0.09(10)  &  +0.01(10)  &  -1.24(19) \\
 3  &  ON  &  ON  &  L $-$ R &  ON  &20.8(16) &\multicolumn{2}{c}{ lead absorber present}& -0.01(19) &  +0.01(10)  &  +0.11(10)  &  -1.12(19)  \\
 4  &  ON  &  ON  &  R $-$ L &  ON  &22.2(14) &\multicolumn{2}{c}{ lead absorber present}& +0.15(18)&  -0.12(10) & -0.23(10)  & -1.41(23)  \\
 5  &  ON  &  ON  &  L $-$ R &  OFF &25.0(13) &\multicolumn{5}{c}{evaluated ratio 501/(444+332) see text} &  -1.19(13)  \\
 6  &  ON  &  OFF &  R $-$ L &  ON  & 57(7)   &-0.41(16)  &  -1.89(73)  &  -0.12(11)  &  -0.06(11)  &  -0.15(11)  &  -1.48(26)  \\
 7  &  OFF &  ON  &  L $-$ R &  OFF &7.6(1)   &\multicolumn{5}{c}{evaluated ratio 501/444+332) see text} &  -0.93(13)   \\
 8  &  OFF &  ON  &  R $-$ L &  ON  &$>$1000  &-0.11(6)  &  -0.33(25)  &  +0.14(8)  &  -0.10(8) &  -0.13(8) &  +0.11(18) \\ \hline
\multicolumn{6}{l|}{Weighted average asymmetry for}\\ 
\multicolumn{6}{l|}{transitions other than the 501 keV} &  -0.08(9)  &  -0.74(74) &   -0.04(7)  &   -0.07(5)   &   -0.03(8)\\
\multicolumn{6}{l|}{for cold reversals 1 $-$ 7.}\\
\end{tabular}
\end{ruledtabular}
\end{table}
\endgroup 

\section{\label{further}Further analysis.}
\subsection{\label{fra}The fraction in good sites: calibration of the $^{\rm 180}$Hf$^{\rm m}$ thermometer.} 
One non-ideal feature of the use of ion implantation for sample preparation is the fact that a fraction of the implanted hafnium nuclei come to rest at sites which do not experience the full, substitutional site, hyperfine interaction. Such sites include not only irregular sites in the iron matrix, but also nuclei which undergo strong interactions in the surface layers of the target and come to rest in the thin oxide layer which is always present on iron foils. The low-temperature nuclear orientation technique does not have the ability to explore details of the site distribution, but investigations have shown that frequently a valid description of the system is to consider a two-site model, with fraction $f$ in the substitutional site and the remainder in a zero field site thereby remaining unoriented at all temperatures. The model can be shown to be valid in a particular case if the fraction $f$ extracted using it is found to be constant, independent of the sample temperature.

The substitutional site hyperfine interaction is known to about 6\%, as discussed above, providing an extremely high degree of polarization of this fraction at the lowest temperatures reached in this work. The angular distribution W$_{\rm cal}$ of the $^{\rm 180}$Hf$^{\rm m}$ gamma transitions from nuclei in such sites can be calculated as a function of temperature from their known multipole character. A value of the fraction $f$ can then be extracted from the data by comparison of the measured anisotropy W$_{\rm exp}$ of each of the 215, 332, 444 and 501 keV transitions, provided the temperature is known, using the relation [W$_{\rm exp}$($\theta$) - 1] = f [W$_{\rm cal}$($\theta$) -  1].

\begin{figure}
\centerline{\psfig{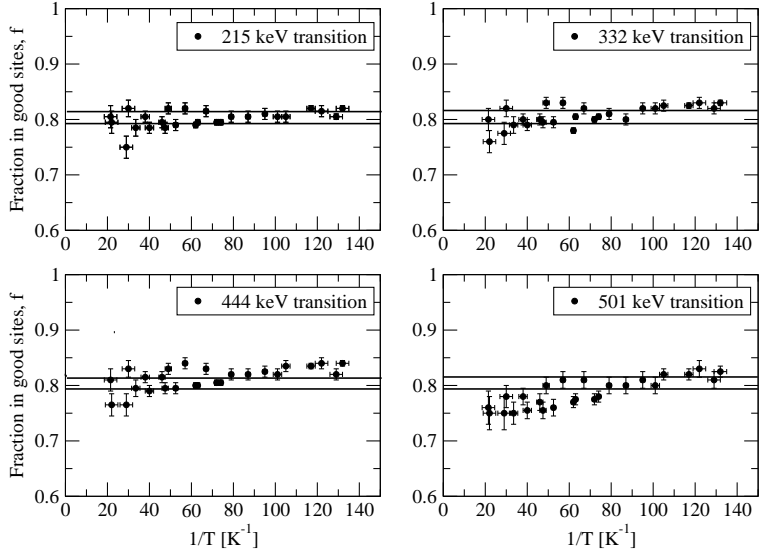}}
\caption{\label{fig6}Fraction in good sites, $f$, evaluated from data on $^{\rm 180}$Hf$^{\rm m}$ decay transitions as described in the text. The full lines give the range of the determined value $f$=0.805(10).}
\end{figure}

The sample was cooled from above 50 mK to the base temperature of 7.6 mK twice during the experiment, the initial and final cool-downs, both times with temperatures determined from the observed anisotropies of the pure electric quadrupole 137 keV transition of $^{\rm 57}$Co. The $^{\rm 180}$Hf$^{\rm m}$ data on all gamma transitions from all files taken during both cool-downs and at base temperature have been analysed for the fraction $f$. The results are shown in Fig.~\ref{fig6}. It is seen that a value of $f$ close to 80\% is obtained from each transition, over the full temperature range of the experiment, with only a slight upward drift (1-2\%), justifying use of the simple two-site distribution model. The value adopted for further analysis of the data is $f$  =  0.805(10), as indicated in Fig.~\ref{fig6}.

With the fraction $f$ determined, measured anisotropies of the $^{\rm 180}$Hf$^{\rm m}$ gamma transitions can be used as thermometers. This was done in analyzing data of reversals 2,3,4, and 6, when the presence of lead absorbers removed the $^{\rm 57}$Co transitions from the spectra.

\subsection{\label{temp}Temperatures of reversals 5 and 7.}
These two reversals were carried out without the pulser peak present in the gamma spectra. This means that whilst relative anisotropies and asymmetries of different gamma transitions can be evaluated, absolute, source strength corrected, values for individual transitions are not available and thus temperature was not directly measured. However indirect arguments can be made to establish the temperature for these reversals. 

\begin{figure}
\centerline{\psfig{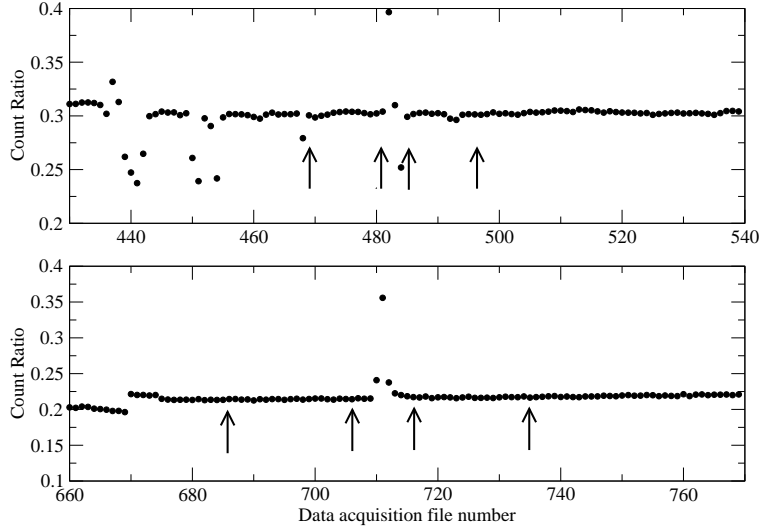}}
\caption{\label{fig7}Upper panel: Ratio of counts in the 444 keV gamma transition peak in one axial (0$^{\rm o}$, 180$^{\rm o}$) detector and the 90$^{\rm o}$ detector for the time period spanning reversal 5. The generally constant level of this ratio shows constancy of temperature and determines the temperature of reversal 5. The variations seen between files $\sim$ 430 and 470 are caused by interruption in the hafnium beam, whilst those around file 485 are due to the field reversal itself. The ranges of files used for pre-reversal and post-reversal averages are indicated by vertical arrows.
Lower panel: As in the upper panel but for the period spanning reversal 7, showing the data used to determine the temperature of reversal 7 (see text). The step at file 670 was produced by dead-time changes when the pulser was removed. The statistical errors are smaller than the data point symbols.}
\end{figure}

$^{\rm 180}$Hf$^{\rm m}$ transitions with pulser present were available for thermometry until shortly before reversal 5 and after it the pulser was reconnected and lead absorbers removed so that both $^{\rm 180}$Hf$^{\rm m}$ and $^{\rm 57}$Co thermometry were available. At this stage of the experiment the implanted beam was steady with time and the sample activity had reached its asymptotic value, thus there was no reason to expect the sample temperature to vary. Fig.~\ref{fig7} (upper panel) shows the ratio of the number of counts in the 444 keV transition peak recorded in one axial detector to the number recorded in the 90$^{\rm o}$ detector during the period containing the reversal. The figure shows constancy of this ratio, that is constancy of the sample temperature, to within about one millikelvin over this period. $^{\rm 57}$Co thermometry shortly after the field reversal (files 540-548) gave the temperature as 25.0(13) mK, which is assigned to reversal 5.

Reversal 7 was performed about 6 hours after the beam was stopped. The dilution refrigerator cooled sharply, reaching base temperature at least 3 hours before the reversal. The 444 keV un-normalized 0$^{\rm o}$/90$^{\rm o}$ ratio as above is shown in Fig.~\ref{fig7} (lower panel), 
 again indicating constant temperature after the cool-down over the period of the reversal. The $^{\rm 57}$Co temperature measured at the end of the cool-down was 7.6(1) mK, which is taken as the temperature for reversal 7.

The `dummy' reversal 8 was performed after warm-up of the source to above 1 K and analysed as the other reversals; as required the average asymmetry was zero within statistical errors for all transitions (see Table~\ref{tab1}).
\begin{table} 
 \caption{\label{tab2}Angular distribution coefficients used in calculation of the even polynomial terms 
in the anisotropy of transitions from  $^{\rm 180}$Hf$^{\rm m}$.}
\begin{ruledtabular}
\begin{tabular}{ccccccc}
 E$_\gamma$ [keV]  &  I$_{\rm i}^\pi$  &  I$_{\rm f}^\pi$  &  Multipolarity  &U$_{\rm 2}$A$_{\rm 2}$Q$_{\rm 2}$  & U$_{\rm 4}$A$_{\rm 4}$Q$_{\rm 4}$  & U$_{\rm 6}$A$_{\rm 6}$Q$_{\rm 6}$\\
\hline       
   501$^*$	&   8$^{\rm -}$ &  6$^{\rm +}$  &   M2/E3   &   -0.407(11)   &   +0.106(9)   &   -0.222(8)\\
   444  &   8$^{\rm +}$ &  6$^{\rm +}$  &   E2      &   -0.338(1)   &   -0.114(1)   &    0.000\\
   332  &   6$^{\rm +}$ &  4$^{\rm +}$  &   E2      &   -0.332(1)   &   -0.107(1)   &    0.000\\
   215	&   4$^{\rm +}$ &  2$^{\rm +}$  &   E2      &   -0.332(1)   &   -0.107(1)   &    0.000\\
 \hline
\multicolumn{4}{l}{$^*$ For this transition all U$_\lambda$ $\equiv$ 1.}
\end{tabular}
\end{ruledtabular}

\end{table}
\subsection{\label{ext}The parity non-conserving E2/M2 multipole mixing ratio of the 501 keV transition in $^{\rm 180}$Hf$^{\rm m}$.}
The asymmetry, $\mathcal{A}$, as defined by Eq.~\ref{asymmetry}, was evaluated for each field reversal (except 5 and 7) for all transitions observed (the $^{\rm 57}$Co transitions were not observed when the lead absorbers were present) with the results given in Table~\ref{tab1}.
They show that there is a clearly established non-zero result of between 0.9\% and 1.5\% for the 501 keV transition in every reversal in which the nuclear sample was polarized. All other transitions show zero asymmetry within statistical error, and there is no evidence of asymmetry outside error in these and the 501 keV transition in the `dummy' reversal 8. The larger experimental error for the 137 keV transition is caused by its lower intensity and poorer peak-to-background ratio than the other transitions (see Fig.~\ref{fig3}). 

 Accepting the evidence that the 332 keV and 444 keV transitions show no effect, data for the 5th and 7th reversals were analysed using the ratio of the 501 peak counts to the sum of the 332 and 444 keV peak counts. It is shown in Appendix A that this ratio can be used to calculate the asymmetry of the 501 keV transition on the assumption that the other transitions have zero asymmetry.

The asymmetries measured for the 501 keV transition are given in Table~\ref{tab3} and plotted versus 1/T in Fig.~\ref{fig8} 
where they may be compared with theoretical calculations of the asymmetry obtained, using the same hyperfine interaction and fraction in good sites as mentioned above, for selected values of the E2/M2 mixing ratio $\epsilon$. The weighted average value of this mixing ratio is $\epsilon$ = - 0.0324(16) (Table~\ref{tab3}).

\subsection{\label{error}Error analysis.}

The uncertainties in this experiment are of three types; statistical, theoretical and geometrical. The result given at the end of the previous section shows only the statistical error associated with the spectral data. This is simple to estimate as windows were set on the photopeaks, with no peak fitting, and background was taken from adjacent windows above and below the well-resolved peaks.. The other two sources of uncertainty are discussed in this section, leading to the final result. A second possible source of error in the 501 keV photopeak count is pile-up of pulses from detection of coincidences, either true or accidental, between 444 keV and 57 keV quanta. This effect has been estimated by comparing the weak pile-up peaks observed for other energy pairs (215 keV + 332 keV etc) with the product of their individual photopeak counts. This established that the contribution of pile-up events in the 501 photopeak was 0.4(1)\% of the true single 501 quantum counts when the lead absorbers were absent, and orders of magnitude less when they were in place. Since, on field reversal, the change of total count rate is of order 1\%, hence a maximum 1\% change of the pile-up rate itself, pile-up has negligible influence on the measured asymmetry of the 501 kev transition.

\begin{figure}
\centerline{\psfig{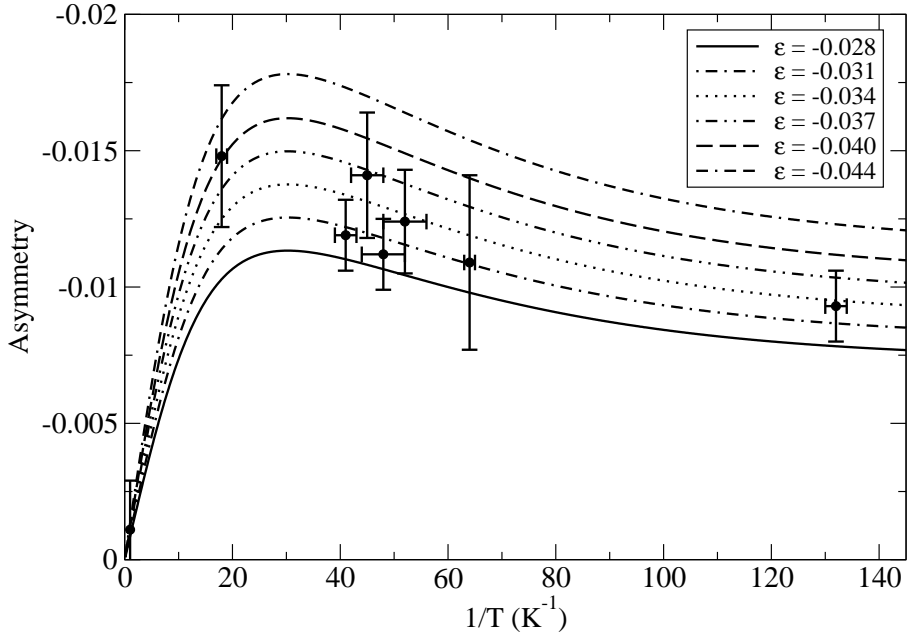}}
\caption{\label{fig8}Measured asymmetry $\mathcal{A}$ of the 501 keV transition, as a function of inverse temperature, compared with calculations using Eq.~\ref{asymmetry} for a range of values of the E2/M2 mixing ratio $\epsilon$.}
\end{figure}

The theoretical uncertainty derives from the adopted values of three parameters; the hyperfine interaction strength taken to be 25.9(13) mK, the fraction in good sites $f$ measured to be 0.805(10) and the ¡normal¢ E3/M2 mixing ratio $\delta$ in the 501 keV transition, taken to be 5.3(3). To estimate the consequence of the uncertainties in these parameters on the value of the E2/M2 mixing ratio calculations of the asymmetry $\mathcal{A}$ as a function of $\epsilon$ and inverse temperature, as in Fig.~\ref{fig8} were made using extreme values and compared with the ¡standard¢ calculations in which the central values, 25.9 mK, 0.805 and 5.3 were taken. The results of these calculations showed that uncertainty in the interaction, the fraction $f$ and the E3/M2 mixing ratio produced changes in values of $\epsilon$ deduced from the measured asymmetries by, respectively, $\pm$ 1.9\%, $\pm$ 2.7\% and $\pm$ 4.9\%. Added in quadrature, the total uncertainty, from these causes, in the average value of $\epsilon$ is $\pm$ 0.0017.

The geometrical uncertainty concerns the accuracy in positioning of the detectors, e.g their angles to the orientation axis, and in estimation of their solid angle correction factors. Allowing for a range of $\pm$ 2$^{\rm o}$ in angle and of $\pm$ 1\% in the solid angle correction factors produced changes in extracted $\epsilon$ of about 1\%, but in opposite directions. These small and cancelling corrections have been omitted from the final error calculation.  The high degree of cancellation between the $\lambda$ = 1 term, which is negative and the $\lambda$ = 3, 5 terms which are positive, in the asymmetry calculation, renders the result less sensitive to the solid angle corrections than might be expected and also insensitive to temperature except at very low values of 1/T.

The final result is therefore $\epsilon$ = - 0.0324(16)(17) where the first uncertainty is based on measurement statistics and the second the result of uncertain input parameters. 
 
\begin{table} 
\caption{\label{tab3}Results of asymmetry $\mathcal{A}$(T) in the 501 keV transition for all reversals, in order of decreasing temperature, and the extracted E2/M2 mixing ratio $\epsilon$.}
 \begin{ruledtabular}
\begin{tabular}{crrrr}
Reversal No. &  T[mK] &  1/T[K$^{\rm -1}$] & $\mathcal{A}$[\%] & $\epsilon$[\%]\\ \hline
  6  &  57(7)    &  17(2) &  -1.48(26)  &  -3.8(7) \\
  5  &  25.0(13)  &  40(2) &  -1.19(13)  &  -3.0(3) \\
  4  &  22.2(14)  &  45(3) &  -1.41(23)  &  -3.7(6) \\
  3  &  20.8(16)  &  48(3) &  -1.12(13)  &  -3.0(5) \\
  2  &  19.2(13)  &  52(4) &  -1.24(19)  &  -3.4(5) \\
  1  &  15.6(2)  &  64(1) &  -1.09(32)  &  -3.1(9) \\
  7  &  7.6(1)   &  132(2) &  -0.93(13)  &  -3.4(5) \\ \hline  
\multicolumn{4}{l}{Average $\epsilon$ = - 3.24(16)\%} \\ 
\hline
\end{tabular}
\end{ruledtabular}
\end{table}

\section{\label{disc}Discussion.}

The measurements were designed to eliminate several possible sources of systematic error. As detailed in Table~\ref{tab1}, reversals were carried out from different initial field directions, counting rates were varied by the addition of absorbers and reversals were done with and without the pulser, the presence of which gave rise to some peak distortion in the gamma spectra. The excellent stability of the beam spot position on the iron foil is demonstrated by the steadiness of the 332 keV/444 keV L/R ratio (shown in Fig.~\ref{fig4}) throughout the experiment and also by the close equality of the `warm' ratios taken at the start and finish of the experiment. The temperature was varied as widely as possible to give considerable change in the degree of polarization of the hafnium nuclei, thus providing the broadest possible evidence for the temperature variation of the asymmetry. The results show the required robust self-consistency under all conditions.
The final result for the irregular E2/M2 mixing ratio in the 501 keV transition, $\epsilon$ = - 0.0324(16)(17), is in extremely close agreement with the long accepted best value -0.030(2)\cite{kra75}. Thus the present work, done with up-to-date and very different technology from the previous measurements, endorses fully the older results and has extended them. The observed temperature variation of the asymmetry, taken to significantly higher degrees of nuclear polarization than was accessible previously, shows full agreement with the behaviour predicted by the earlier result for $\epsilon$. This is a very satisfactory outcome which, if it lacks excitement, preserves the status of the asymmetry of this transition as the best established demonstration of parity admixture in nuclear phenomena. Theory thus faces the problem of providing understanding of the numerical result. Although the present experiment was not in every way fully optimized, there is little incentive to go further experimentally until some theoretical stimulus arises.

Unfortunately, to date the problems presented by the specific nuclear structure of the levels involved, notably the high degree of K-forbiddeness of the 8$^{\rm -}$ decay, have prevented effective theoretical calculation of the expected degree of parity non-conservation in this case. Thus the very factor which is in all likelihood responsible for the large magnitude of the effect, the strong hindrance of the normal decay matrix elements, is also the cause of difficulty in giving a good theoretical description of the observed large parity violating effect.

\section{Acknowledgements}
\label{sec5}
We wish to thank S.~Chamoli and B.~S.~Nara Singh for assistance with data taking during the experiment. Support from the ISOLDE and NICOLE collaborations is gratefully acknowledged.
This research was sponsored by the US DOE grants DE-FG02-96ER40983 (UT) and  DE-FG02-94ER40834 (UMD), EPSRC (UK), the Fund for Scientific Research Flanders (FWO), the European Union Sixth Framework through R113-EURONS (contract no. 506065), EU-RTD project TARGISOL (HPRI-CT-2001-50033), the Israel Science Foundation and the The Ministry of Education of the Czech Republic 1P04LA211 and Istitutional Research Plan AV0Z10480505.

\section{\label{appA}Appendix A}
For reversals 5 and 7 normalization of the 501 keV transition to the pulser was not available and the ratio of the intensity of this transition to the sum of the combined 444 keV and 332 keV transitions was used as a measure of the 0$^{\rm o}$--180$^{\rm o}$ asymmetry. Here it is shown that this is equivalent to the pulser normalized asymmetry provided the other transitions contain negligible parity violating terms in their angular distributions.

For each transition and for a single detector at angle $\theta$ to the axis of orientation, the recorded gamma count in any file, N$_{\rm i}$ can be written as the sum of product terms as in Eq.~\ref{w}, but it is convenient to separate the even, parity conserving terms from the odd, parity non-conserving terms, thus
\begin{equation}
\mathcal{S} f_{\rm i} \mathcal{E}_{\rm i} W(\theta,E_{\rm \gamma}^{\rm  i})  =  \mathcal{S} c_{\rm i}(1+e_{\rm i}+o_{\rm i})
\label{a1}
\end{equation}
where e(o)$_{\rm i}$  = $\sum\limits_{\rm k \, \rm even(odd)} B_\lambda U_\lambda A_\lambda Q_\lambda P_\lambda(\cos\theta)$, $\mathcal{S}$ is the instantaneous source strength, c$_{\rm i}$ is the product of the fractional gamma emission per decay f$_{\rm i}$ and the detector efficiency $\mathcal{E}_{\rm i}$ for that energy.

The ratio $R$ of counts in the 501 keV peak N$_{\rm 501}$ to the sum of counts in the 444 keV peak, N$_{\rm 444}$, and the 332 keV peak, N$_{\rm 332}$, measured in the detector in the direction of the applied field $\theta$ = 0$^{\rm o}$(180$^{\rm o}$) pre(post) reversal, can be written:
\begin{eqnarray}
R_{\rm pre(post)}(\theta)&=&\left(\frac{N_{\rm 501}}{N_{\rm 332} + N_{\rm 444}}\right)_{\rm pre(post)} \nonumber\\
 & =& \frac{\mathcal{S}c_{\rm 501}(1 + e_{\rm 501}(\theta)  +  o_{\rm 501}(\theta))}{\mathcal{S}c_{\rm 332}(1 + e_{\rm 332}(\theta)  + o_{\rm 332}(\theta))+\mathcal{S}c_{\rm 444}(1 + e_{\rm 444}(\theta)  + o_{\rm 444}(\theta))}.
\end{eqnarray}
The difference between the ratios $R$ measured in the initially 0$^{\rm o}$ detector pre and post reversal, divided by their sum, after cancellation of the source strength, is given by
\begin{equation}
\mathcal{R} = \frac{R_{\rm pre}-R_{\rm post}}{R_{\rm pre}+R_{\rm post}}
\end{equation}
where the change of detector position from $\theta$ = 0$^{\rm o}$ (pre-reversal) to $\theta$ = 180$^{\rm o}$ (post-reversal) reverses the sign of the odd terms but does not change the even terms.

After straightforward but somewhat lengthy manipulation this ratio becomes
\begin{equation}
\mathcal{R} = 
\frac{
o_{\rm 501}[(c_{\rm 332}(1 + e_{\rm 332}) + c_{\rm 444}(1 + e_{\rm 444})]  -  c_{\rm 332}o_{\rm 332}(1 + e_{\rm 501})  - c_{\rm 444}o_{\rm 444}(1 + e_{\rm 501})}
{c_{\rm 332}(1 + e_{\rm 332})(1 + e_{\rm 501}) + c_{\rm 444}(1 + e_{\rm 444})(1 + e_{\rm 501}) - o_{\rm 501}(c_{\rm 332}o_{\rm 332} + c_{\rm 444}o_{\rm 444})}.
\end{equation}
When the product of odd terms (that is of parity non-conserving terms in the 501 keV transition and in either of the other transitions) in the denominator is neglected as small, this further simplifies to
\begin{equation}
\mathcal{R}=\frac{ o_{\rm 501}}
{(1 + e_{\rm 501})}    
- \frac
{c_{\rm 332}o_{\rm 332} + c_{\rm 444}o_{\rm 444}}
{c_{\rm 332}(1 + e_{\rm 332}) + c_{\rm 444}(1 + e_{\rm 444})}
\end{equation}
which can be seen to be half the asymmetry of the 501 keV transition minus half the weighted sum of the the asymmetries of the 332 keV and 444 keV transitions.

If it is assumed that the other asymmetries are indeed zero, then the ratio $\mathcal{R}$ is simply half the asymmetry of the 501 keV transition. This assumption is adopted in the analysis of reversals 5 and 7, both because the experiment provides evidence for the other asymmetries being small and because of the theoretical difference between the strong intra-band nature of the 332 keV and 444 keV E2 transitions and the highly K-forbidden 501 keV M2/E3 transition.

In reversals 5 and 7 the two axial detectors yield independent values of $\mathcal{R}$, $\mathcal{R}$(0$^{\rm o}$) and $\mathcal{R}$(180$^{\rm o}$), of opposite signs, since one detector is initially at 0$^{\rm o}$ and the other at 180$^{\rm o}$. The combination $\mathcal{R}$(0$^{\rm o}$) - $\mathcal{R}$(180$^{\rm o}$) (equals 2 $\mathcal{R}$(0$^{\rm o}$)) gives the full asymmetry of the 501 keV transition. This is given in Tables~\ref{tab1} and \ref{tab3}.


\newpage
\begin{thebibliography}{99}
\bibitem{jen70}
B.~ Jenschke and P.~Bock, Phys. Lett. {\bf{31B}}, 65 (1970)
\bibitem{lip71}
E.~D.~Lipson, P.~Boehm and J.~C.~Vanderleeden, Phys. Lett. {\bf{35B}}, 307 (1971)
\bibitem{kra71a}
K.~S.~Krane, C.~E.~Olsen, J.~R.~Sites and W.~A.~Steyert, Phys. Rev. Lett. {\bf{26}}, 1579 (1971).
\bibitem{kra71b}
K.~S.~Krane, C.~E.~Olsen, J.~R.~Sites and W.~A.~Steyert, Phys. Rev. {\bf{C4}}, 1906 (1971).
\bibitem{kra72}
K.~S.~Krane, C.~E.~Olsen and W.~A.~Steyert, Phys. Rev. {\bf{C5}}, 1663 (1972).
\bibitem{kra75}
T.~S.~Chou, K.~S.~Krane and D.~A.~Shirley, Phys. Rev. {\bf{C12}}, 286 (1975).
\bibitem{kra86}
K.~S.~Krane in \textit{Low Temperature Nuclear Orientation}, eds. N.~J.~Stone and H.~Postma, North Holland Amsterdam, Chapter 6, (1986).
\bibitem{kup74}
E.~Kuphal, P.~Dewes and E.~Kankeleit, Nucl. Phys. {\bf{A234}}, 308 (1974).
\bibitem{lip72}
E.~D.~Lipson, P.~Boehm and J.~C.~Vanderleeden, Phys. Rev. {\bf{C5}}, 932 (1972).
\bibitem{nar05}
B.~S.~Nara Singh et al., Phys. Rev. {\bf{C72}}, 027303 (2005).
\bibitem{ade85}
E.~C.~Adelberger and W.~C.~Haxton, Ann. Rev. Nucl. Part. Sci. {\bf{35}}, 501 (1985).
\bibitem{ede90}
R.~Eder et al., Hyp. Int. {\bf{59}}, 83 (1990).
\bibitem{kos07}
U.~K\"oster et al., in print in Eur. Phys. J. A. 
\bibitem{mar86}
H.~Marshak, in \textit{Low Temperature Nuclear Orientation}, eds. N.~J.~Stone and H.~Postma, North Holland Amsterdam, Chapter 16, (1986).
\bibitem{wal80}
P.~M.~Walker, D.~Ward, O.~H\"ausser, H.~R.~Andrews and T.~Faestermann, Nucl.Phys. {\bf{A349}}, 1 (1980)
\bibitem{sto05}
N.~J.~Stone, Atomic Data and Nuclear Data Tables {\bf{90}}, 75 (2005). 
\bibitem{rao}
G.~N.~Rao, Hyp.Int. {\bf{24/26}}, 119 (1985).
\bibitem{nie02}
A.~Nieminen et al., Phys. Rev. Lett. {\bf{88}} 094801 (2002). 
\bibitem{mut04}
S.~Muto, T.~Ohtsubo, S.~Ohya and K.~Nishimura, Hyp. Int. {\bf{158}}, 195 (2004).
\bibitem{kor71}
H.~J.~K\"{o}rner, F.~E.~Wagner and B.~D.~Dunlap, Phy. Rev. Lett. {\bf{37}}, 1593 (1971).
\bibitem{rei67}
P.~G.~E.~Reid, M.~Shott and N.~J.~Stone, Phys.Lett.A {\bf{25}}, 456 (1967).
\bibitem{shaw}
T.~Shaw and N.~J.~Stone, Atomic Data and Nuclear Data Tables {\bf{42}}, 339 (1989).
\bibitem{lau74}
R.~Laurenz. E.~Klein and W.~D.~Brewer, Z. Physik {\bf{A270}}, 233 (1974).
\bibitem{wu03}
S-C.~Wu and H.~Niu, Nucl. Data Sheets {\bf{100}}, 483 (2003).
\end{thebibliography}
\end{document}